\documentclass[citeautoscript,prb,reprint,superscriptaddress,noshowpacs]{revtex4-1}
\usepackage{amsmath,graphicx,xcolor,siunitx}
\usepackage[utf8]{inputenc}
\newcommand{\gzrgm}{\hbox{$\Gamma$--Z--R--$\Gamma$--M}}
\newcommand{\gzr}{\hbox{$\Gamma$--Z--R}}
\newcommand{\gr}{\hbox{$\Gamma$--R}}
\newcommand{\rgm}{\hbox{R--$\Gamma$--M}}
\newcommand{\zrg}{\hbox{Z--R--$\Gamma$}}

\newcommand{\Jsaf}{\ensuremath{J_\mathrm{s}^\mathrm{AF}}}
\newcommand{\Jsfm}{\ensuremath{J_\mathrm{s}^\mathrm{FM}}}
\newcommand{\Jwaf}{\ensuremath{J_\mathrm{w}^\mathrm{AF}}}
\newcommand{\Jwfm}{\ensuremath{J_\mathrm{w}^\mathrm{FM}}}
\newcommand{\Joo}{\ensuremath{J_\mathrm{o..o}}}

\newcommand{\cuseo}{\ensuremath{\text{Cu}_2\text{OSeO}_3}}

\begin{document}
\title{Spin excitations in the skyrmion host \cuseo{}}
\author{G~S~Tucker}
\affiliation{Laboratory for Quantum Magnetism, École Polytechnique Fédérale de Lausanne, CH-1015 Lausanne, Switzerland}
\affiliation{\hbox{Laboratory for Neutron Scattering and Imaging,~Paul Scherrer Institut, CH-5232 Villigen, Switzerland}}
\author{J~S~White}
\affiliation{\hbox{Laboratory for Neutron Scattering and Imaging,~Paul Scherrer Institut, CH-5232 Villigen, Switzerland}}
\author{J~Romhányi}
\affiliation{Institute for Theoretical Solid State Physics, IFW Dresden, D-01069 Dresden, Germany}
\author{D~Szaller}
\author{I~Kézsmárki}
\affiliation{Department of Physics, Budapest University of Technology and Economics and MTA-BME Lendület Magneto-optical Spectroscopy Research Group, 1111 Budapest, Hungary}
\author{B~Roessli}
\author{U~Stuhr}
\affiliation{\hbox{Laboratory for Neutron Scattering and Imaging,~Paul Scherrer Institut, CH-5232 Villigen, Switzerland}}
\author{A~Magrez}
\affiliation{Crystal Growth Facility, École Polytechnique Fédérale de Lausanne, CH-1015 Lausanne, Switzerland}
\author{F~Groitl}
\affiliation{Laboratory for Quantum Magnetism, École Polytechnique Fédérale de Lausanne, CH-1015 Lausanne, Switzerland}
\affiliation{\hbox{Laboratory for Neutron Scattering and Imaging,~Paul Scherrer Institut, CH-5232 Villigen, Switzerland}}
\author{P~Babkevich}
\author{P~Huang}
\author{I~Živković}
\author{H~M~Rønnow}
\affiliation{Laboratory for Quantum Magnetism, École Polytechnique Fédérale de Lausanne, CH-1015 Lausanne, Switzerland}

\begin{abstract}
We have used inelastic neutron scattering to measure the magnetic excitation spectrum along the high-symmetry directions of the first Brillouin zone of the magnetic skyrmion hosting compound \cuseo{}. 
The majority of our scattering data are consistent with the expectations of a recently proposed model for the magnetic excitations in \cuseo{}, and we report best-fit parameters for the dominant exchange interactions. 
Important differences exist, however, between our experimental findings and the model expectations.
These include the identification of two energy scales that likely arise due to neglected anisotropic interactions. 
This feature of our work suggests that anisotropy should be considered in future theoretical work aimed at the full microscopic understanding of the emergence of the skyrmion state in this material.
\end{abstract}
\maketitle

Magnetic skyrmions are topologically non-trivial spin structures that can extend over tens of nanometers \cite{bogdanov_thermodynamically_1989,bogdanov_thermodynamically_1994,nagaosa_topological_2013}.
In certain magnetic compounds with non-centrosymmetric crystal structure they can condense 
and form a regular hexagonal arrangement
as observed in the metallic helimagnets MnSi \cite{muhlbauer_skyrmion_2009}, $\text{Fe}_{1-x}\text{Co}_x\text{Si}$ \cite{munzer_skyrmion_2010}, FeGe \cite{yu_near_2011}, and CoZnMn \cite{tokunaga_new_2015}, insulating \cuseo{} \cite{adams_long-wavelength_2012}, and in the polar magnetic semiconductor $\text{GaV}_4\text{S}_8$ \cite{kezsmarki_neel-type_2015}.
To understand the formation and the microscopic origin of these skyrmion phases one needs a multi-scale approach that covers the macroscopic domain of the skyrmion as well as the quantum scale of the local spins. 
This however breaks down in the above mentioned metals, because the low energy delocalized electrons and magnetic degrees of freedom are mixed, intrinsically involving multiple energy and spatial scales.

Among cubic helimagnets \cuseo{} is the only insulator with magnetoelectric properties in the ground state \cite{seki_observation_2012,adams_long-wavelength_2012,seki_magnetoelectric_2012,white_electric-field-induced_2014,white_electric_2012,omrani_exploration_2014}.
It offers an ideal laboratory to explore the microscopic ingredients that lead to skyrmion formation 
in a quantitative manner, since its Bloch-type ground state properties and low energy excitations are fully governed by the magnetic interactions between localized spins and are not affected by the presence of itinerant carriers. 
Exchange pathway considerations, susceptibility measurements, and \textit{ab initio} calculations reveal that two magnetic energy scales divide the system into weakly coupled $\text{Cu}_4$ tetrahedra \cite{janson_quantum_2014}. 
These $\text{Cu}_4$ ``molecules'', with an effective spin of $S=1$, are the elementary magnetic building blocks of \cuseo{} instead of the single Cu ions. 
The effective spins of the $\text{Cu}_4$ tetrahedra are ferromagnetically coupled and form a trillium lattice just as the Mn and Fe ions do in the B20 structure of the metallic skyrmion compounds MnSi and FeGe.

Prior to the undertaking of the present work, previous studies of the magnetic excitation spectra of \cuseo{} were conducted using Raman scattering \cite{gnezdilov_magnetoelectricity_2010} and microwave resonance absorption \cite{kobets_microwave_2010}; techniques that are sensitive only to excitations in the center of the Brillouin zone. 
In contrast, inelastic neutron scattering (INS) is able to measure at finite momentum transfer and is therefore uniquely suited to probe the magnetic excitation spectra of \cuseo{} throughout reciprocal space. 
The additional information afforded by INS therefore provides more rigorous tests of theoretical models aimed at describing the excitation spectra of \cuseo{}.

Single crystals of \cuseo{} (cubic $P2_13$ space group, $a= 8.82$ \AA) were grown via chemical vapor transport as described elsewhere \cite{miller_magnetodielectric_2010,belesi_ferrimagnetism_2010}. 
Three \cuseo{} single crystals of $\sim 1$ g total mass were coaligned with $[110]$ and $[001]$ in the horizontal scattering plane. 
The magnetic properties of each individual crystal were verified by magnetization measurements, and subsequent neutron diffraction confirmed that the mosaic sample displayed a transition temperature between magnetically ordered and disordered states at $T_\mathrm{c}=57.1(6)$~K, consistent with previous reports \cite{bos_magnetoelectric_2008,belesi_ferrimagnetism_2010}. 
INS measurements were performed at the thermal triple-axis neutron spectrometer EIGER and the cold triple-axis neutron spectrometer TASP, both located at the Swiss Spallation Neutron Source (SINQ), Paul Scherrer Institut, Switzerland. 
The sample mosaic was installed into a standard Orange cryostat which provided a base temperature of $1.5$~K. 
Inelastic scans were performed in constant-$k_\mathrm{f}$ mode, with $k_\mathrm{f}=2.662$ and $4.1$ \AA$^{-1}$ at EIGER and $1.55$ \AA$^{-1}$ at TASP, for $\mathbf{q}$ points along the line \gzrgm{} around several $\Gamma$ points.

\begin{figure}
\includegraphics{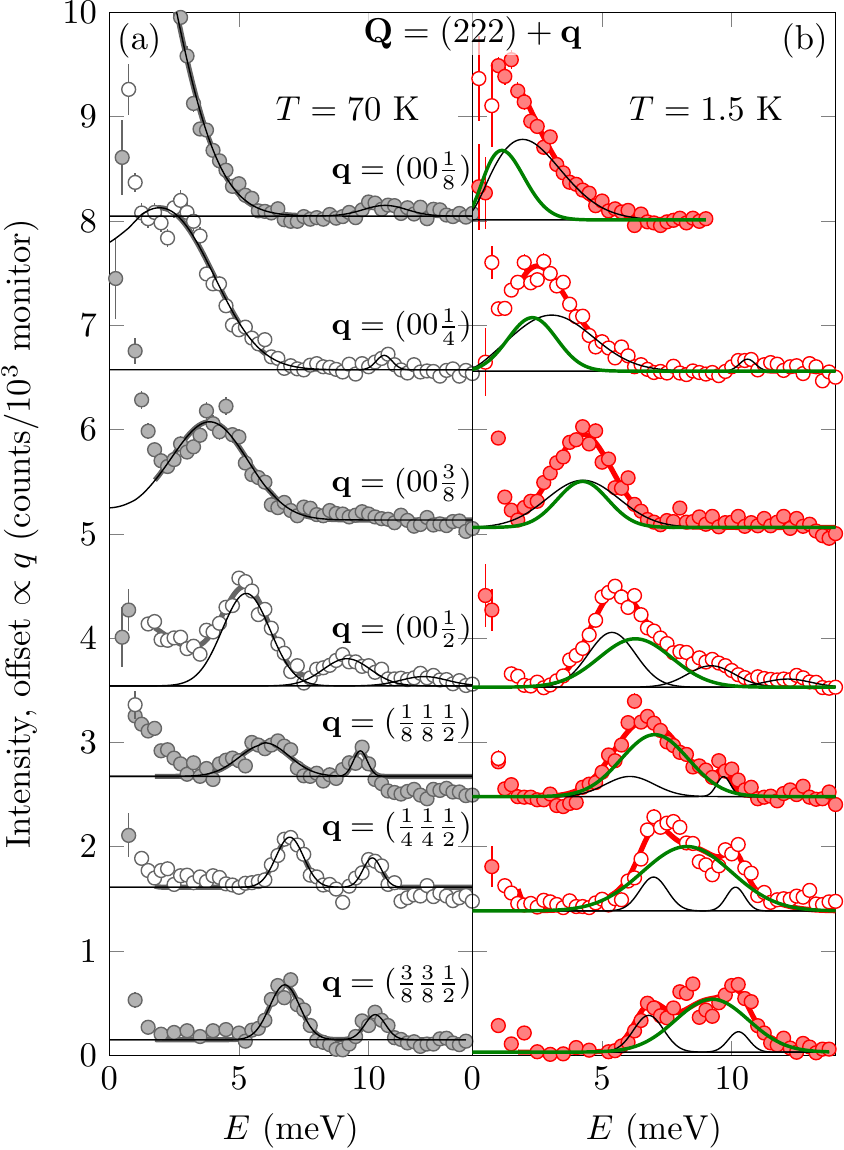}
\caption{
INS intensity (circles) measured at EIGER for constant $\mathbf{Q}$ as a function of energy transfer, along the line \gzr{} [$(222)$--$(22\frac{5}{2})$--$(\frac{5}{2}\frac{5}{2}\frac{5}{2})$].
Scans in panel (a) were measured at $T=70$ K, above $T_\mathrm{c}$, where all peaks are the result of lattice excitations (phonons); thick solid lines represent the best-fit to a function comprised of a number of temperature-dependent Gaussian peaks, thin lines represent the individual phonon peaks in each scan.
Scans in panel (b) were measured at $T=1.5$ K and have been fit to the same function used to describe the $70$ K data (with phonon parameters fixed and phonon intensity rescaled to account for thermal population effects) plus an additional Gaussian to account for magnetic inelastic scattering; thick and thin lines are as in panel (a), medium thickness (green) lines represent the magnetic scattering peaks.
Adjacent scans are offset by an amount proportional to $q$.
\label{fig:disp222GZR}}
\end{figure}

Figure \ref{fig:disp222GZR} shows representative INS data collected at EIGER for a series of constant-$\mathbf{Q}$ scans performed along the reciprocal space line \gzr{} around $(222)$. 
Data in Fig.~\ref{fig:disp222GZR}(a) were collected in the paramagnetic state at $T=70$~K where the excitation spectra at the probed energy scale is devoid of peaked magnetic scattering and is dominated by lattice excitations (phonons). 
To capture the phonon intensity, individual scans from this high-temperature data were fit by one or more peaks, consisting of a Gaussian multiplied by the Bose thermal factor. 
Data in Fig.~\ref{fig:disp222GZR}(b) were collected at our base temperature of $T=1.5$~K and at the same $\mathbf{Q}$ points as in panel (a).
They contain a peaked magnetic response in addition to the phonon scattering. 
The low-temperature data were fit by combining the high-temperature phonon model (with all peak parameters fixed) plus an additional Gaussian peak to capture the magnetic scattering.
By comparing the $\mathbf{q}$-dependence of the phonon and magnetic excitation peak positions it is clear that the two have different dispersion relations, thus confirming the different physical origins of the high and low temperature INS intensities.
\footnote{The magnetic excitation dispersion can also be obtained simply by taking the Bose-factor-corrected difference between the low- and high-temperature datasets -- we confirmed that qualitatively similar results are obtained for the dispersion relation via this method.}

\begin{figure}
\includegraphics{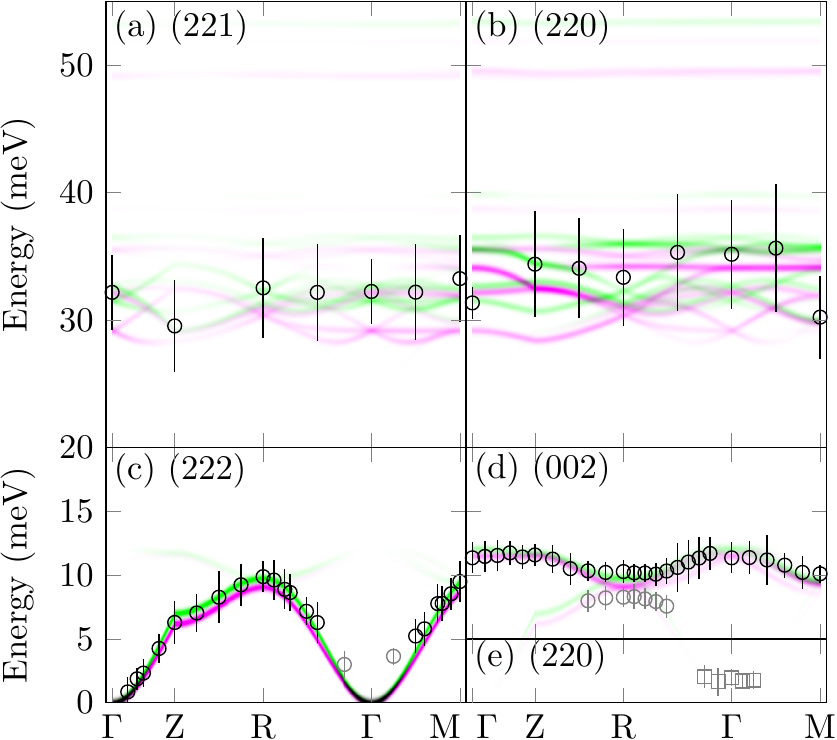}
\caption{
Magnetic excitation dispersion as measured on EIGER (circles) and TASP (squares) overlaid on model intensity calculations for the set of exchange parameters reported in Ref.~\onlinecite{romhanyi_entangled_2014} (magenta) and our best-fit exchange parameters (green), both detailed in table \ref{tab:params}.
Panels (a) and (b) show the high-energy dispersion around $(221)$ and $(220)$, respectively.
Panels (c), (d), and (e) show the low-energy dispersion around $(222)$, $(002)$, and $(220)$, respectively.
All points represent fit peak-positions from constant $\mathbf{Q}$ energy scans with vertical bars indicating the full-width-at-half-maximum of each fit peak, which are mainly dominated by the instrumental resolution.
Black points were included in our fitting routine while gray points were excluded.
\label{fig:dispmap}}
\end{figure}

Figure \ref{fig:dispmap} shows the magnetic dispersion obtained from our INS data along the \gzrgm{} line around four $\Gamma$-points. 
In addition, the figure shows a comparison between the measured dispersion and two calculated neutron scattering intensity maps. 
The experimental data points in Fig.~\ref{fig:dispmap} display vertical bars that are indicative of the measured peak width arising from the finite energy resolution of the instrument.
One intensity map is that expected according to the set of exchange parameters proposed in Ref.~\onlinecite{romhanyi_entangled_2014}, the other is our best-global-fit set of exchange parameters. 
The two parameter sets produce qualitatively similar intensity maps with our best-global-fit solution producing a better quantitative result.

\begin{figure}
\includegraphics{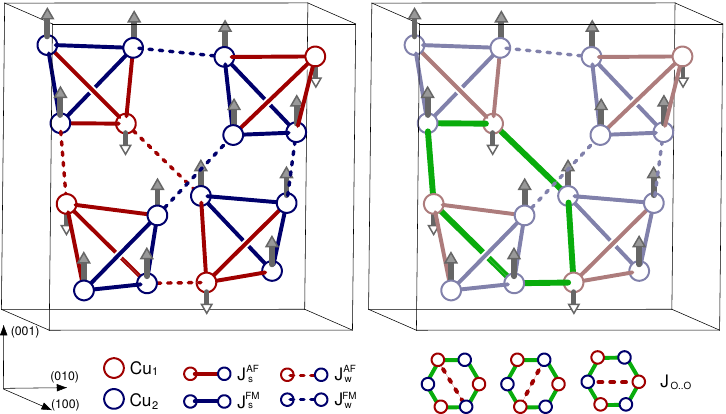}
\caption{%
A sketch of the magnetic unit cell of \cuseo{}. 
The unit cell contains 16 $\text{Cu}^{2+}$ ions located in two symmetry inequivalent sites.
The $\text{Cu}_1$ and $\text{Cu}_2$ sites for a network of coupled tetrahedra and are respectively represented by red and blue circles.
Four of the five model exchange couplings defined in the text are indicated on the left.
The fifth, $\Joo{}$ couples opposite $\text{Cu}_1$ and $\text{Cu}_2$ sites across the alternating $\text{Cu}_1$--$\text{Cu}_2$ hexagon in the unit cell, and is indicated on the right.
\label{fig:exchanges}}
\end{figure}
Next we introduce the theoretical model against which we test our experimental data. 
In Ref.~\onlinecite{romhanyi_entangled_2014}, the excitation spectra of \cuseo{} is calculated within the framework of a multiboson formalism for the constituent $\text{Cu}_4$ tetrahedra that includes five Heisenberg-like exchange interactions, indicated schematically in Fig.~\ref{fig:exchanges}. 
The two strongest exchange parameters, \Jsaf{} and \Jsfm{}, couple the spins within a single $\text{Cu}_4$ tetrahedra. 
Two weaker exchange parameters couple the spins between $\text{Cu}_4$ tetrahedra, \Jwaf{} and \Jwfm{}, and a final parameter, \Joo{}, couples across alternating $\text{Cu}_1$--$\text{Cu}_2$ hexagons \cite{romhanyi_entangled_2014}.

By comparing our measured dispersion with the model calculations, a clear sensitivity to the energy-scale and bandwidth of the low-energy acoustic and optical magnetic modes is found that determines the relationship between the three weakest couplings. 
Although we are unable to resolve the details of the high-energy dispersion expected according to the model, 
our measurements also prove to be sensitive to the energy-scale and overall-bandwidth of the modes at higher-energies, which fix the relationship between \Jsaf{} and \Jsfm{}. 
\begin{figure}
\includegraphics{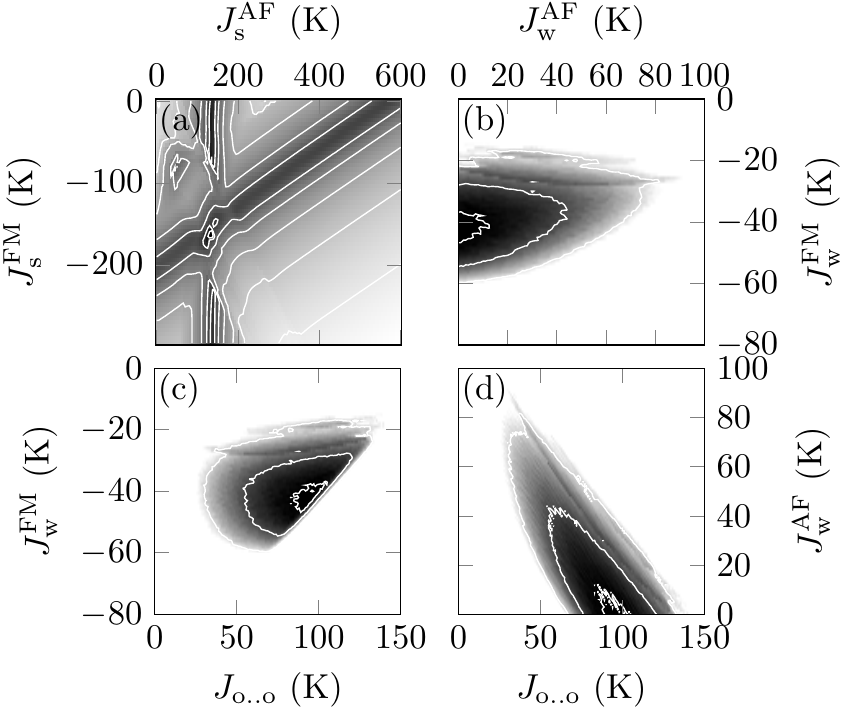}
\caption{%
Sum of the squared difference in energy (SSE) between measured and calculated peak positions as a function of (a) strong or (b,c,d) weak exchange parameters.
The SSE for a point is encoded in its shade, with black indicating small SSE and white indicating large SSE. Overlaid with the SSE maps are constant-SSE contour lines.
The map in panel (a) is the SSE calculated from only the high-energy dispersion at the indicated (\Jsaf{},\Jsfm{}) points.
The maps in panels (b,c,d) are minimum-value projections of the SSE calculated from only the low-energy dispersion at points in a (\Jwaf{},\Jwfm{},\Joo{}) grid. 
For the indicated values of \Jwaf{} and \Jwfm{}, panel (b) shows the minimum SSE independent of \Joo{}; similarly panels (c) and (d) show the minimum SSE independent of \Jwaf{} and \Jwfm{}, respectively.
\label{fig:gridSSE}}
\end{figure}
By computing the sum of the squared difference in energy (SSE) between our data, field-dependent electron spin resonance data \cite{ozerov_establishing_2014}, Raman data \cite{gnezdilov_magnetoelectricity_2010}, and far-infrared data \cite{miller_magnetodielectric_2010} and the model-calculated dispersion on two independent grids throughout 2D (\Jsfm{},\Jsaf)- and 3D (\Jwfm{},\Jwaf{},\Joo{})-parameter space, we found a single minimum in weak-parameter space and many local minima in strong-parameter space, as shown in Fig.~\ref{fig:gridSSE}.
By starting a Levenberg-Marquardt least-squares fitting routine near the various minima in five-dimensional parameter space and comparing best-local-fit SSE as well as full predicted spectra, we have found a set of best-global-fit parameters which are detailed in table \ref{tab:params}.
\begin{table}
\caption{
Model exchange parameters used to produce the intensity maps displayed in Fig.~\ref{fig:dispmap}, with noted color for each set of parameters.
Positive coupling values correspond to antiferromagnetic interactions while negative values are ferromagnetic.
Standard deviations of the best-global-fit parameters are given in parentheses in units of the last digit.
\label{tab:params}}
\begin{tabular}{S[table-format=3.1]S[table-format=4.1]S[table-format=2.2]S[table-format=3.1]S[table-format=3.1]cc}\toprule
{\Jsaf{}/K} & {\Jsfm{}/K} & {\Jwaf{}}/K & {\Jwfm{}/K} & {\Joo{}/K} & Color & Reference \\\hline
145 & -140   & 28 & -50    &  45    & magenta & \onlinecite{romhanyi_entangled_2014}\\
%133.6808\pm4.6280 &-157.2864\pm6.7828 & 4.8414\pm0.4600 & -42.0174\pm5.2010 & 90.6255\pm7.8847 & green   & this work\\
135\pm5 & -157\pm7 & 4.8\pm0.5 & -42\pm5 & 91\pm8 & green   & this work\\\toprule
\end{tabular}
\end{table}

\begin{figure}
\includegraphics{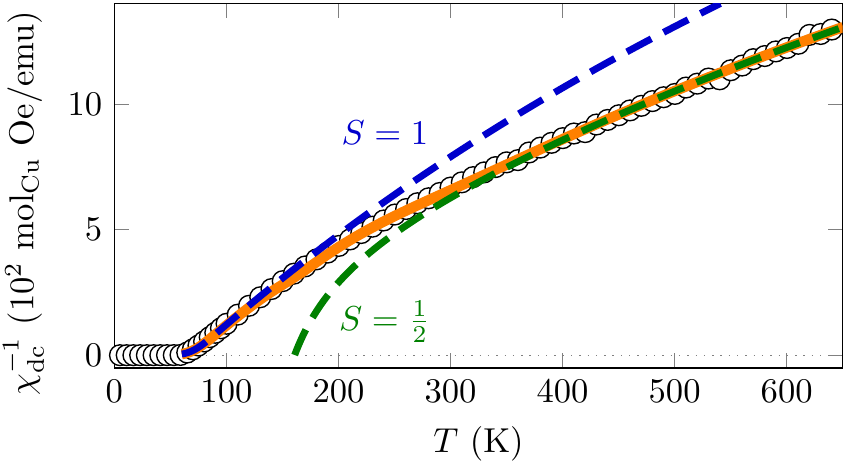}
\caption{%
Inverse dc magnetic susceptibility data (symbols) from Ref.~\onlinecite{zivkovic_two-step_2012}.
Dashed lines represent the $S$=$1$ (low-temperature) and $S$=$1/2$ (high-temperature) limits of the model magnetic susceptibility, as described in the text.
The solid line -- given by $\chi(T)=\sum_S\chi_S\{1-(-1)^{2S} \tanh [(T-t)/w ]\}/2$ where $S=\frac{1}{2},1$ -- smoothly transitions between the two limits via coincident antisymmetric step functions with step parameters fit to $t$=$242(2)$ K and $w$=$78(2)$ K.
\label{fig:invchi}}
\end{figure}
A mean-field approximation for the high-temperature susceptibility of this model gives
\begin{equation}
\chi_\frac{1}{2}=\frac{C_\frac{1}{2}\left[T-\frac{3}{8}(\Jsaf{}+\Jwaf{}+\Joo{})+\frac{1}{8}(\Jsfm{}+\Jwfm{})\right]}{(T-T_\frac{1}{2})\left[T+T_\frac{1}{2}+\frac{1}{2}(\Jsfm{}+\Jwfm{})\right]},
\label{eq:chioh}
\end{equation}
with $C_\frac{1}{2}=N_\mathrm{A} g^2 \mu_\mathrm{B}^2 S(S+1)/3k_\mathrm{B}$, $S=\tfrac{1}{2}$, $g$=$2$, and 
\begin{equation}
\begin{split}
T_\frac{1}{2}= &\frac{1}{4}\sqrt{(\Jsfm{}+\Jwfm{})^2+3(\Jsaf{}+\Jwaf{}+\Joo{})^2} \\
     &-\frac{1}{4}(\Jsfm{}+\Jwfm{}).
\end{split}
\label{eq:tcoh}
\end{equation}
At low temperatures, the strong interactions prevail and each strong tetrahedra behaves as a single $S=1$ spin which gives $\chi_1=C_1/T-T_1$ with $C_1=(N_\mathrm{A}/4) g^2 \mu_\mathrm{B}^2 S(S+1)/3k_\mathrm{B}$, and $T_1=5(\Jwaf{}-5\Jwfm{}/3+\Joo{})/12$.
The high- and low-temperature approximations for the magnetic susceptibility allow for a direct comparison of the model and our best-global-fit parameters to published magnetic susceptibility data with only a single fitting parameter $\chi_0$ via $\chi(T)=\chi_S+\chi_0$.
Using our best-global-fit exchange parameters and $\chi_0=1.844(14)\times10^{-4}$ emu/mol$_\mathrm{Cu}$ Oe, Fig.~\ref{fig:invchi} shows good agreement between inverse magnetic susceptibility data from Ref.~\onlinecite{zivkovic_two-step_2012} and the high-  and low-temperature approximations for the susceptibility.

\begin{figure}
\includegraphics{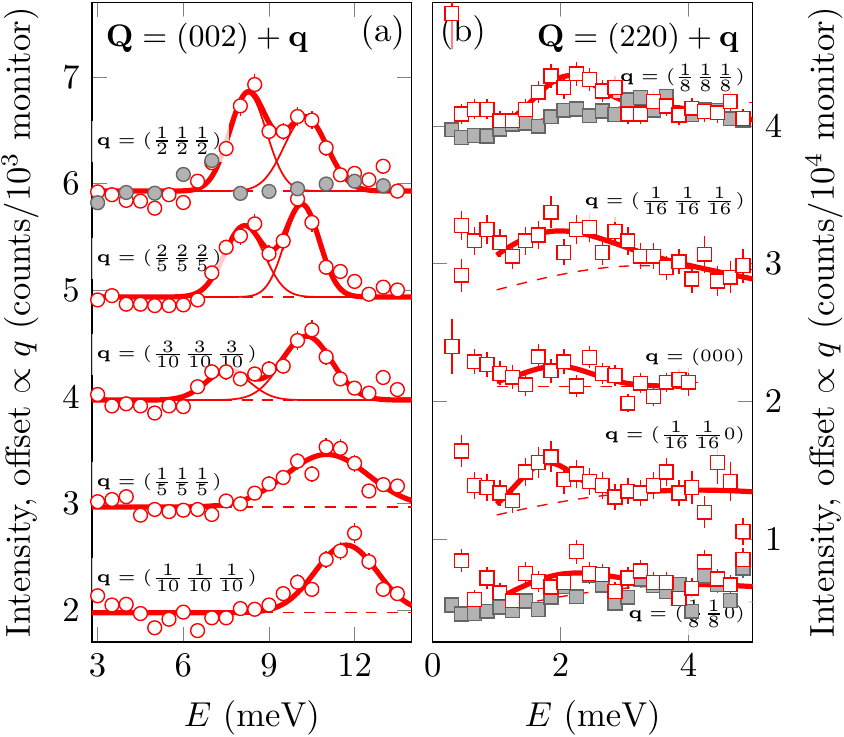}
\caption{
Representative scans showing features not predicted by the model in Ref.~\onlinecite{romhanyi_entangled_2014}.
Panel (a), scans along the line \gr{} [$(002)$--$(\frac{1}{2}\frac{1}{2}\frac{5}{2})$] performed on EIGER (circles) with $k_\mathrm{f}=2.662$ \AA$^{-1}$.
Panel (b), scans along \rgm{} [$(\frac{5}{2}\frac{5}{2}\frac{1}{2})$--$(220)$--$(\frac{5}{2}\frac{5}{2}0)$] performed on TASP (squares) with $k_\mathrm{f}=1.55$ \AA$^{-1}$.
Open symbols are constant-$\mathbf{Q}$ energy scans performed at the indicated $\mathbf{Q}$ points and $T=1.5$ K;
filled symbols are data measured at $T=60$ (circles) or $70$ K (squares) rescaled by the ratio of their Bose thermal population factors and that at $1.5$ K;
solid lines are fits to the $1.5$ K data and dashed lines are an estimate of the non-magnetic background.
\label{fig:disp002GZ}}
\end{figure}

Finally we discuss aspects of our experimental data that depart qualitatively from the theoretical expectations of Ref.~\onlinecite{romhanyi_entangled_2014}. 
In the dispersion of the magnon modes we observed two features which are not predicted by the model.
As shown in Fig.~\ref{fig:dispmap}(d), the first is a $\sim 2$ meV splitting along the line \zrg{}. 
This splitting is between the acoustic and optic intertetrahedral modes, and is shown in closer detail in Fig.~\ref{fig:disp002GZ}(a). 
The second deviation between our data and the model expectation is the observation of a seemingly broad and weakly dispersive, low-energy excitation at $\sim2$ meV near $\Gamma$, as shown in Fig.~\ref{fig:dispmap}(e), and in detail in Fig.~\ref{fig:disp002GZ}(b).
We find that no set of parameters can coax the hitherto applied model to reproduce these two features seen in our data.
Due to the fact that zone-center measurements show the dispersion at $\Gamma$ to have a gap no larger than $\sim12$ $\mu$eV \cite{kobets_microwave_2010}, any single-ion anisotropy is likely to be small and we expect instead that these unexplained features are related to the network of antisymmetric, i.e., Dzyaloshinskii-Moriya (DM), interactions in this material \cite{yang_strong_2012}. 
These chiral interactions are ultimately responsible for the stabilization of the slightly incommensurate helical groundstate and field-induced skyrmion phases \cite{adams_long-wavelength_2012,seki_formation_2012,white_electric-field-induced_2014}.
Using our data to fit an extended model including anisotropy will hence allow quantification of these pivotal DM interactions.
Related to this, the associated helimagnon excitations are expected to be closely-spaced, and located at low energy close to $\Gamma$ \cite{janoschek_helimagnon_2010,kugler_band_2015}. 
Within our finite energy resolution, the presence of these excitations could contribute to the low energy feature in our data, though the energy-scale of the helimagnon bands is not expected to extend up to $\sim2$ meV in \cuseo{} \cite{janoschek_helimagnon_2010}. 
Further spectroscopy experiments with improved energy resolution are needed to unveil the nature of these low energy excitations.

Through inelastic neutron scattering experiments we have shown that the magnetic excitation spectrum of \cuseo{} exhibits an overall agreement with a proposed model that makes use of five Heisenberg-like exchange parameters to describe the coupling between the 16 $\text{Cu}^{2+}$ ions in the unit cell.
By comparing INS peak positions with those expected according to model calculations, we have restricted the five-dimensional parameter space to a single best-fit point that differs from those previously proposed\cite{romhanyi_entangled_2014}.
Our dataset also reveals two energy scales that are not expected in theory; the splitting of the optical magnetic excitation near $(002)$, and a weakly dispersive feature at low energy near $(220)$. 
We propose that these features could arise due to antisymmetric interactions neglected by the model. 
%The presence of these features highlights the importance of including anisotropic effects in future attempts to understand fully the magnetic excitations spectrum, and ultimately the microscopic description of the nanometric length-scale skyrmion lattice state.
The presence of these features suggests that anisotropic effects should be considered in future attempts to fully understand the magnetic excitation spectrum, and ultimately the microscopic description, of the nanometric length-scale skyrmionic spin texture.

\emph{Note Added:}~During the preparation of this paper we became aware of another neutron spectroscopy report \cite{portnichenko_magnon_2015}. 
The data in that report are in overall agreement with ours, but the splitting of the magnetic excitation near R is not reported.

\begin{acknowledgments}
Neutron scattering experiments were performed at the Swiss Spallation Neutron Source (SINQ), Paul Scherrer Institut, Switzerland. 
Financial support from the Swiss National Science Foundation, the European Research Council grant CONQUEST and MaNEP is gratefully acknowledged. 
I.~Ž.~acknowledges financial support from the Croatian Science Foundation, Project No. 02.05/33.
I.~K.~and D.~Sz.~were supported by the Hungarian Research Fund OTKA K 108918.
\end{acknowledgments}

\end{document}